\documentclass[conference,a4paper,final,twocolumn,10pt,twoside]{IEEEtran}
\usepackage{layout}
\IEEEoverridecommandlockouts 

\ifCLASSINFOpdf

\else
\fi

\usepackage{mathtools}
\usepackage{amsmath}
\usepackage{amssymb}
\usepackage{cite}
\usepackage{graphicx}
\usepackage{soul}
\usepackage{siunitx}

\begin{document}
%
\title{Mining DNA Sequences Based on Spatially Coded  Technique Using Spatial Light Modulator}
\author{\IEEEauthorblockN{Masoome Fazelian, Sajjad AbdollahRamezani, Sima Bahrani, Ata Chizari, Mohammad Vahid Jamali,\\ Pirazh Khorramshahi, Arvin Tashakori, Shadi Shahsavari, and Jawad A. Salehi,~\IEEEmembership{Fellow,~IEEE}}
\IEEEauthorblockA{Optical Networks Research Laboratoy (ONRL), Sharif University of Technology (SUT), Tehran, IRAN\\
Corresponding Author Email: jasalehi@sharif.edu\\
}}

\maketitle
\begin{abstract}

In this paper, we present an optical computing method for string data alignment applicable to genome information analysis. By applying moir{\'e} technique to spatial encoding patterns of deoxyribonucleic acid (DNA) sequences, association information of the genome and the expressed phenotypes could more effectively be extracted. Such moir{\'e} fringes reveal occurrence of matching, deletion and insertion between DNA sequences providing useful visualized information for prediction of gene function and classification of species. 
Furthermore, by applying  a cylindrical lens, a new technique is proposed to map two-dimensional (2D) association information to a one-dimensional (1D) column of pixels, where each pixel in the column is representative of superposition of all bright and dark pixels in the corresponding row. By such a time-consuming preprocessing, local similarities between two intended patterns can readily be found by just using a 1D array of photodetectors and post-processing could be performed on specified parts in the initial 2D pattern. We also evaluate our proposed circular  encoding adapted for poor data alignment condition.
Our simulation results together with experimental implementation verify the effectiveness of our dynamic proposed methods which significantly improve system parameters such as processing gain and signal to noise ratio (SNR).
\end{abstract}
\begin{IEEEkeywords}
String data alignment, moir{\'e} pattern, DNA sequencing, spatial light modulator.
\end{IEEEkeywords}
\section{Introduction}
Emerging various widespread human diseases speeds up the growing rate of genomics. Accordingly, analysis of deoxyribonucleic acid (DNA) sequences, as a medium storing by far more important information about properties of an organism, has intrigued many researchers to extract significant knowledge about life sciences \cite{kinser2000mining,shendure2008next, rajan2014two}. As a common event in evolution process, mutation would modify DNA data sequences comprising of a finite number of basic elements known as nucleotides, i.e., adenine (A), cytosine (C), guanine (G), and thymine (T), which are independent of each other. Since each sequence data conceals in a collection of one-dimensional (1D) strings forming a genome, the role of string data alignment or pattern matching against a sequence of genomes is much more critical for comparison and interpretation of DNA-based structures \cite{eid2009real}. Due to rapidly evolving DNA-sequencing, investigating through highly extensive DNA databases to identify occurrence of exchange, deletion, and insertion of specific data, find target DNA strings or newly genes and classify species is becoming a costly and challenging problem for researchers \cite{rothberg2011integrated,min2011fast}.

All recent sequencing technologies, including Roche/454, Illumina, SOLiD and Helicos, are able to produce data of the order of giga base-pairs (Gbp) per machine day \cite{metzker2010sequencing}. However, with the emergence of such enormous quantities of data, even the fast digital electronic devices are not effective enough to align capillary reads \cite{ning2001ssaha,kent2002blat}. Actually, today's electronics technology would not permit us to achieve high rate of analysis in sequence matching and information processing due to the time consuming nature of serial processing \cite{tanida1999string,tanida2000string,rothberg2011integrated}. To keep pace with the throughput of sequencing technologies, many new alignment algorithms have been developed, but demands for faster alignment approaches still exist. As a result, the necessity of finding a novel implementation to provide high performance computational systems is undeniable \cite{mardis2008next,merkling2005sequence}. High data throughput, inherent parallelism, broad bandwidth and less-precise adjustment of optical computing provide highly efficient devices which can process information with high speed and low energy consumption. It is worth mentioning that visible light in optical computing systems realizes information visualization for human operators to more effectively carry out genome analysis. Employing a powerful technique to encode DNA information into an optical image besides optical computing capabilities would definitely guarantee to efficaciously perform genomes analysis \cite{mardis2008next,shendure2008next}. 

While recent implementations were static and relying on printed transparent sheets \cite{tanida2002spatially,niita2001genome}, herein, we theoretically and experimentally present dynamic string data alignment based on a spatially coded moir{\'e} technique \cite{amidror2000theory,gabrielyan2007basics} implemented on spatial light modulators (SLMs) which enables one to investigate useful hiding information in genomes. The remaining of the paper is organized as follows. In Section II, the principle of string data matching using the spatially coded  technique is explained. In Section III, bar and circular patterns as an effective scheme for string data alignment will be discussed. In Section IV, the experimental optical architecture and obtained results will be appeared to verify practical feasibility of our proposed pattern, and Section V concludes the paper.
\section{Principles}
\begin{figure}[t]
	\centering
	\includegraphics[width=2.7in]{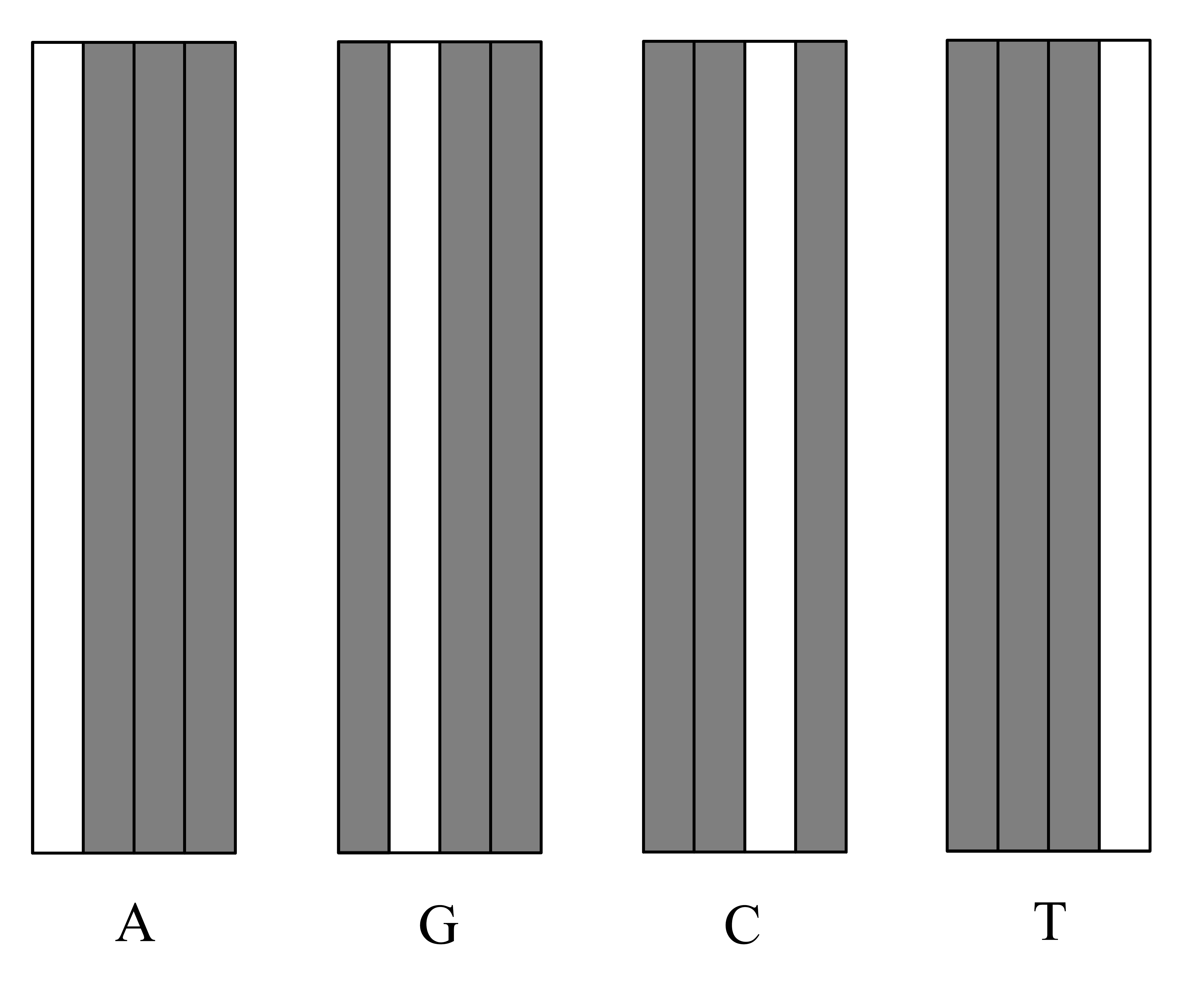}
	\caption{Graphical patterns for DNA bases} 
	\label{graphical}
\end{figure}

In this section, the principles of string alignment by moir{\'e} technique are outlined. Consider two data sequences. The goal of string alignment is evaluation of similarities and differences between them. In particular, we are interested in distinguishing insertion and deletion of elements in any strings with respect to each other. Moir{\'e} technique applies high speed parallel processing of light to perform string alignment. In this approach, four components of strings, namely $\{{\rm A,G,C,T}\}$ are encoded as $\{1000, 0100,0010,0001\}$, respectively. Based on this coding, the strings are spatially coded into images where each component corresponds to four narrow stripes with one bright stripe as $``1"$ and three dark stripes as $``0"$ (see Fig. \ref{graphical}). The coded images are then overlapped with a small relative angle, and by using this technique, correlating segments of the second string in various shifts of the first one can evidently be distinguished. The subsequent matched elements will be appeared as a bright line in the observed pattern of overlapped images.

\begin{figure}[t]
	\centering
	\includegraphics[width=2.7in]{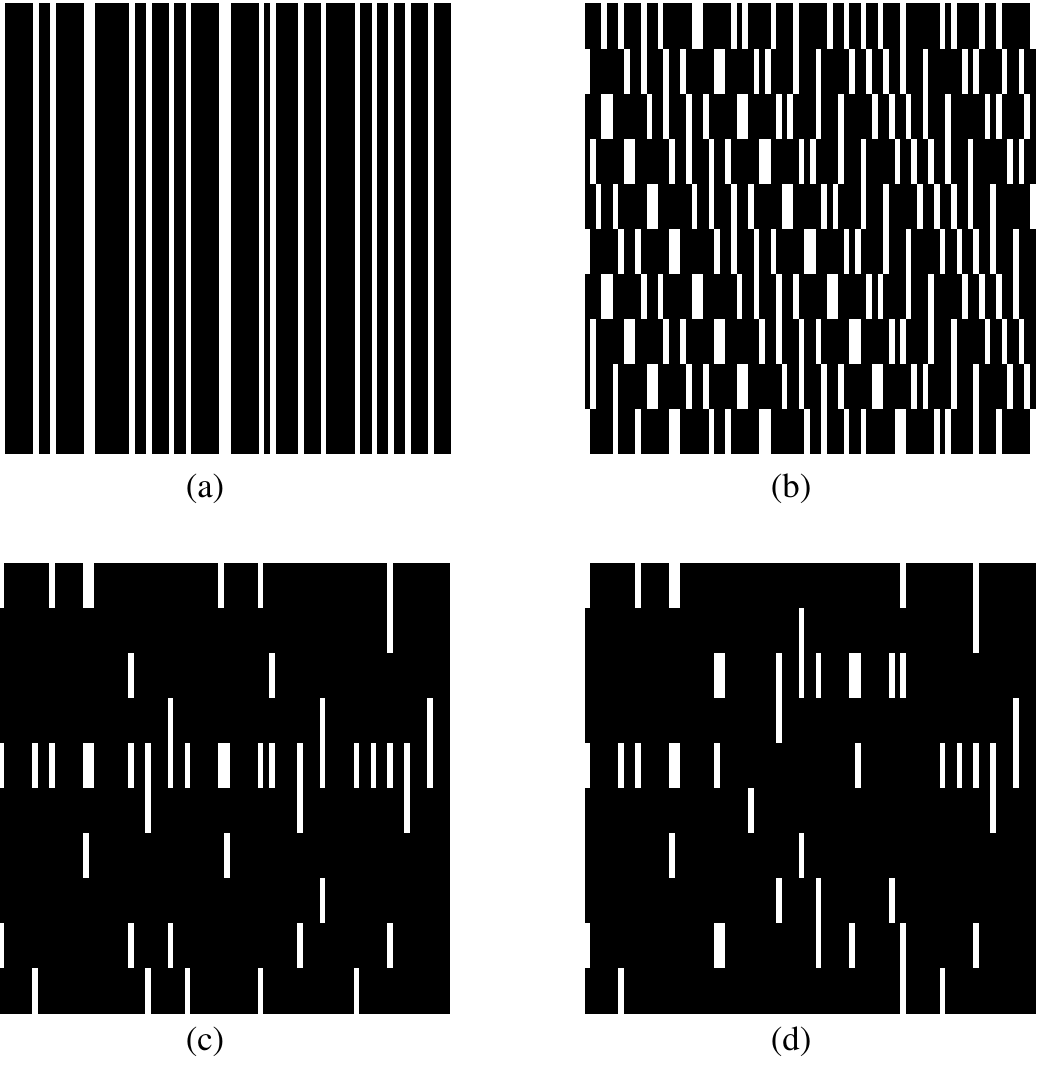}
	\caption{Spatial code patterns for (a) $S_2$, (b) subsequent shifts of initial string $S_1$, (c) output pattern by overlapping (a) and (b), (d) output pattern by overlapping $S_3$ and (b).} 
	\label{fig1}
\end{figure}
As an example, consider two strings S1 of length 40 and S2 of length 20. Now, we want to search for S2 in S1. Fig.~\ref{fig1}(a) shows $S_{2}=\{  {\rm A C G T A T C C G T A C A G G T C G A A} \}$ with respect to the codes appeared in Fig. \ref{graphical}, and each row in Fig.~\ref{fig1}(b) shows subsequent shifts of initial string $S_{1}=\{{\rm T C C G T A C G T A T C C G T A C A G G T C G A A T G C G T A C A T}$ $ {\rm  C G A C C T}\}$; for example first row shows S1(1:20), second row shows S1(2:21), up to the last row. Overlapping Fig.~\ref{fig1}(a) ans (b) results in the pattern shown in Fig.~\ref{fig1}(c); the bright line in the fourth row illustrates that a correlation has happened for a shift of 6, i.e., S2 and S1(6:25) are matched.
\begin{table}[t]
\centering
\caption{Corresponding codes for polarized spatial patterns in Figs. \ref{type1} and \ref{type2}.}
\label{T1}
 \begin{tabular}{c c c c c}  
DNA bases		   & A         &        G &       C &      T  \\ \hline\hline
 
 Type I & $1000$ & $0100$ & $0010$ & $0001$  \\

 Type II & $H00H$ & $V0V0$ & $0V0V$ & $0HH0$ \\ \hline \hline
\end{tabular}
\\

\vspace{.1cm}
\hspace{-3.2cm} $H$: Horizontal, $V$: Vertical
\end{table}
\begin{figure}[t]
	\centering
	\includegraphics[width=3.4in]{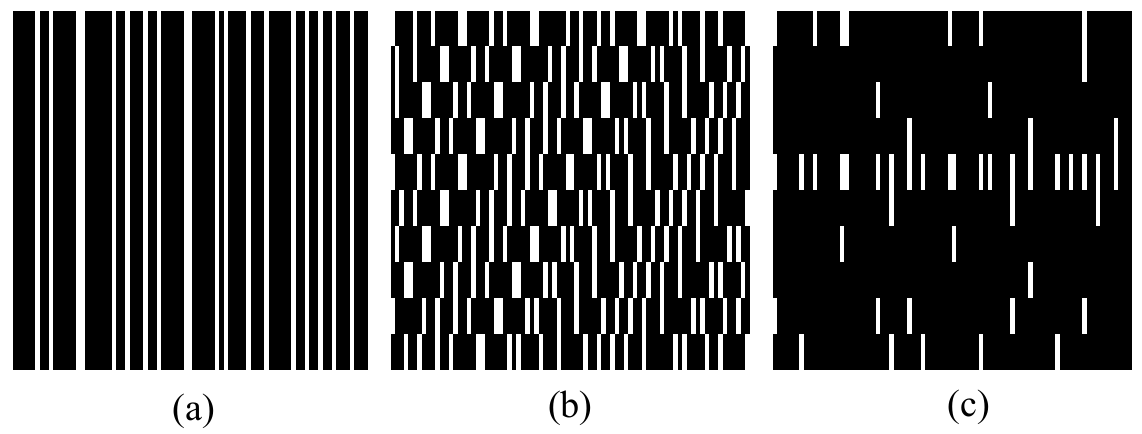}
	\caption {Spatial code patterns of (a) $S_1$, (b) $S_2$, and (c) corresponding correlation for type I.}
	\vspace{0in}\label{type1}
\end{figure}   
\begin{figure}[t]
	\centering
	\includegraphics[width=3.4in]{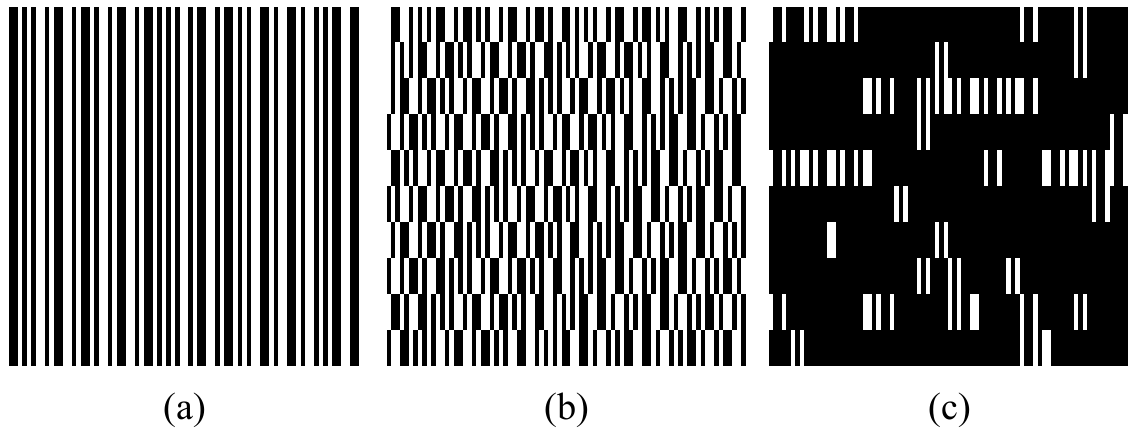}
	\caption {Spatial code patterns of (a) $S_1$, (b) $S_3$, and (c) corresponding correlation for type II.}
	\vspace{0in}\label{type2}
\end{figure}
The insertion and deletion of elements lead to a vertical shift in some parts of the bright line in the overlapping pattern. Each break point indicates the location where insertion or deletion is occurred. The positive and negative vertical shifts correspond to insertion and deletion of some elements, respectively. As an example, consider the string $S_{3}=\{ {\rm A C G T A T\textbf{AG}C C G T A C A}$\st{${\rm G G}$}${\rm T C G A A} \}$ generated by insertion of ``AG'' between the sixth and seventh element of $S_{2}$ and deletion of the fourteenth and fifteenth element of $S_{2}$ . Figure \ref{fig1}(d) depicts the output pattern obtained by multiplying $S_{3}$ and Fig.~\ref{fig1}(b).
\section{Proposed Methods}
In this section, we propose several practically feasible moir{\'e} patterns for string data alignment applications. Wave nature of light provides enough degrees of freedom, i.e., amplitude, phase, and polarization manipulation for sequence data processing.

The first coding approach is based on correlation in which a sequence is simply symbol-by-symbol compared to another sequence. In DNA sequence data processing, each symbol denotes a DNA base. In comparing two symbols with each other, similar symbols generate a bright spot; hence, a correlated  set realizes a bright line. This line is fragmented in the case of insertion and deletion in which vertical distance between fragmented lines identifies the number of deleted or inserted elements  in that place. In this method, SNR can easily be calculated; in the case of two independent and  identical distributed sequences, the probability of such a random similarity and hence number of bright spots with respect to full matching is $0.25$, leading to $6$~dB SNR. It is notable that system's SNR is proportional to the ratio of bright line intensity to the average intensity of other rows.

Another coding technique is based on concatenating two subsequent elements, for example $S(i:i+1)$ and $S(i+1:i+2)$, as a group. Subsequent groups have a common element which ensures an easier detection procedure of insertion or deletion. Coding sequences in overlapped pairs not only does increase the SNR but also makes correlated elements more distinguishable even in the cases of insertion and deletion. In this method, a $12$~dB SNR can be expected in that the probability of random similarity for a word of two symbols is $0.0625$.
\begin{table} [t]
	\centering
	\caption{Corresponding codes for spatial patterns in Figs. \ref{type3} and \ref{type4}.}
	\label{T2}
	\begin{tabular}{c c c}
		
		DNA bases & Type III  & Type IV           \\ \hline \hline
		${\rm AA}$       & $H0000000$& $1000000000000000$\\
		${\rm GA}$	   & $0H000000$& $0100000000000000$\\
		${\rm CA}$	   & $00H00000$& $0010000000000000$\\
		${\rm TA}$	   & $000H0000$& $0001000000000000$\\
		${\rm AG}$       & $0000H000$& $0000100000000000$\\
		${\rm GG}$	   & $00000H00$& $0000010000000000$\\
		${\rm CG}$ 	   & $000000H0$& $0000001000000000$\\
		${\rm TG}$	   & $0000000H$& $0000000100000000$\\
		${\rm AC}$       & $V0000000$& $0000000010000000$\\
		${\rm GC}$	   & $0V000000$& $0000000001000000$\\
		${\rm CC}$	   & $00V00000$& $0000000000100000$\\
		${\rm TC}$	   & $000V0000$& $0000000000010000$\\
		${\rm AT}$       & $0000V000$& $0000000000001000$\\
		${\rm GT}$	   & $00000V00$& $0000000000000100$\\
		${\rm CT}$ 	   & $000000V0$& $0000000000000010$\\
		${\rm TT}$	   & $0000000V$& $0000000000000001$\\ \hline \hline
	\end{tabular}
\end{table}
\begin{table*} 
	\centering
	\caption{Characterization of the proposed method and the corresponding SNRs obtained via simulation, where $N$ stands for the number of SLM pixels. }
	\label{T3}
	\begin{tabular}{c c c c c}
		
		&    Type I    & Type II  & Type III & Type IV \\ \hline \hline
		Processing Gain&	$ N/4$ & $N/4$ & $N/8$ & $N/16$   \\ 
		SLM Modulation Capacity & Intensity or Polarization & Intensity and Polarization & Intensity and Polarization & Intensity or Polarization\\ 
		SNR {$\rm (dB)$} & $6.8854$ & $6.4648$ & $12.2260$ & $12.0715$ \\ \hline \hline
	\end{tabular}
\end{table*}
\subsection{Bar Pattern}
 We examined two different sets of symbols in a bar moir{\'e} pattern. While the first one employs pulse position modulation (PPM), the second comprises of a set of four orthogonal codes using both intensity and polarization (see Table~\ref{T1}). Since there is no useful information in shifts that are not an integer product of symbol length, different rows are shifted by an integer product of four slots that form a symbol. This is by far more efficient than horizontal tilting of the second pattern and consequently compatible with finite resolution of SLM.
 
Simulation results are depicted in  Figs.~\ref{type1} and \ref{type2}. By comparing the results, it is clear that using type II increases the intensity of both noise and signal but does not improve the SNR. Measuring symbol-by-symbol correlation, we see the SNR does not go further than $6$~dB.

\begin{figure}[t]
           \centering
           \includegraphics[width=3.4in]{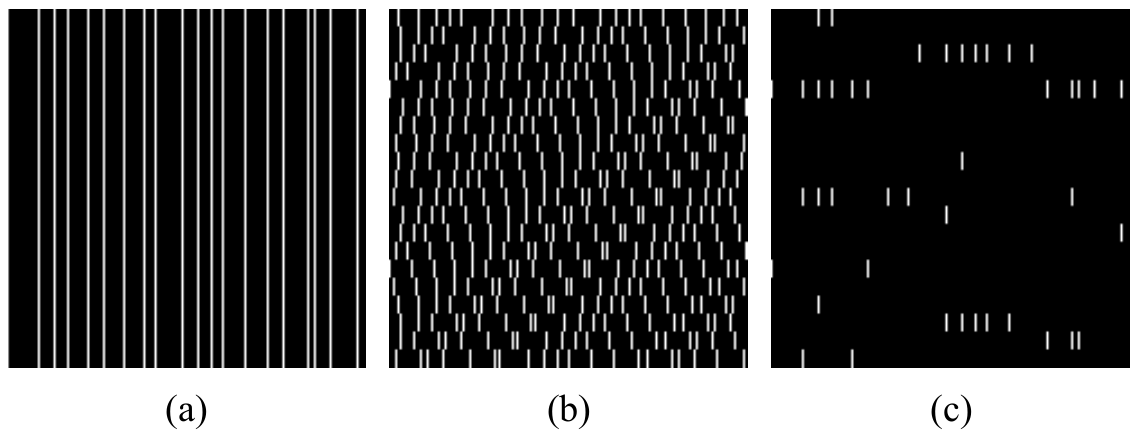}
           \caption {Spatial code patterns of (a) $S_1$, (b) $S_3$, and (c) corresponding correlation for type III.}
                       \vspace{0in}\label{type3}
      \end{figure}

\begin{figure}[t]
           \centering
           \includegraphics[width=3.4in]{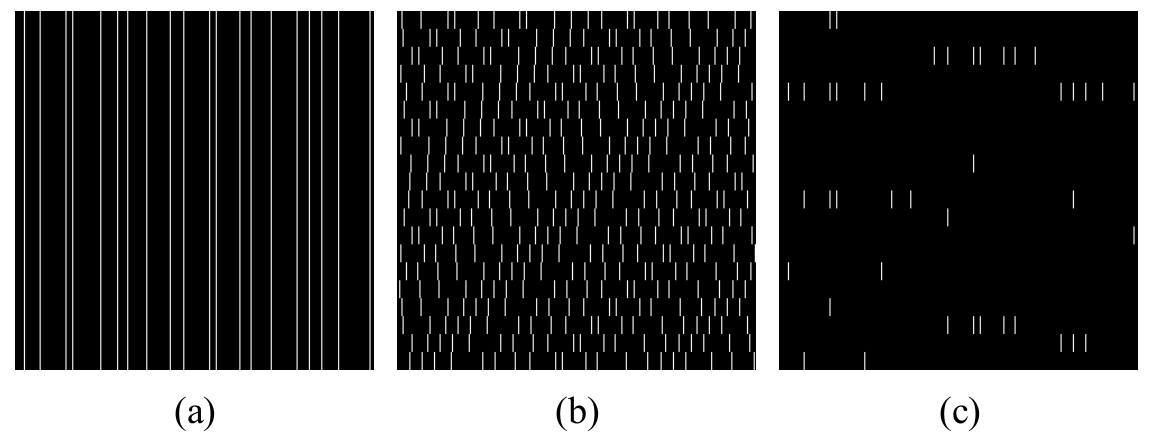}
           \caption {Spatial code patterns of (a) $S_1$, (b) $S_3$, and (c) corresponding correlation for type IV.}
                       \vspace{0in}\label{type4}
      \end{figure}

In the second approach, the codes in Table~\ref{T2} are applied which means that for tilted pattern different rows are shifted by $8k$ in type III (word length is eight here) and $16k$ in type IV; $k$ is a positive integer. Figs.~\ref{type3} and \ref{type4} illustrate the simulation results. As it can be seen, the horizontal straight line is more vivid in types III and IV since the probability of random similarity for a word of two symbols is $0.0625$; therefore, we can expect a SNR about $12$~dB. In type III, we need a SLM with independent intensity and polarization modulation while in type IV only intensity or polarization modulation is required. Polarization modulation can easily be converted to intensity modulation via a polarizer. Since word length for type IV is twice type III, for an equal number of SLM surface pixels, processing gain, the number of DNA bases that the setup is able to compare in each run, of type III is twice type IV. Types III and IV offer better detection capability facing insertion and deletion. It is notable that each insertion or deletion changes two  words. In case of $n$ subsequent deletion or insertion, $n+1$ words differ from initial pattern. Moreover, if we increase the word length to code the DNA bases in group of length $L$, $L-1$ elements are needed to be overlapped in order to detect insertion and deletion.
When misalignment and other types of errors are addressed, the maximum performance of such a system could be achieved. In this case, the number of pixels of SLMs to compare two sequences of length $N$ follows Table \ref{T3}. It also reports the SNR values of different types for a random sequence of length $48$.
\subsection{Circular Pattern}

Optical alignment could be quite problematic in implementing bar patterns. In correlating two bar patterns, the dimension precision required should be about $d/N$, where $d$ is the transverse length of a pixel and $N$ is the total number of vertical pixels on SLM surface.  On the other hand, circular moir{\'e} patterns are basically easier to be adjusted in experimental setups since only the center of circles should be aligned. Besides, it is sensitive to neither rotation nor divergence during propagation in free-space. Despite circular pattern could only process sparse sequence data, it simplifies the optical alignment complexity when transceivers are distant.

\begin{figure}[t]
\centering
\includegraphics[trim=1in 0in 2in 0in ,clip,width=1.8in]{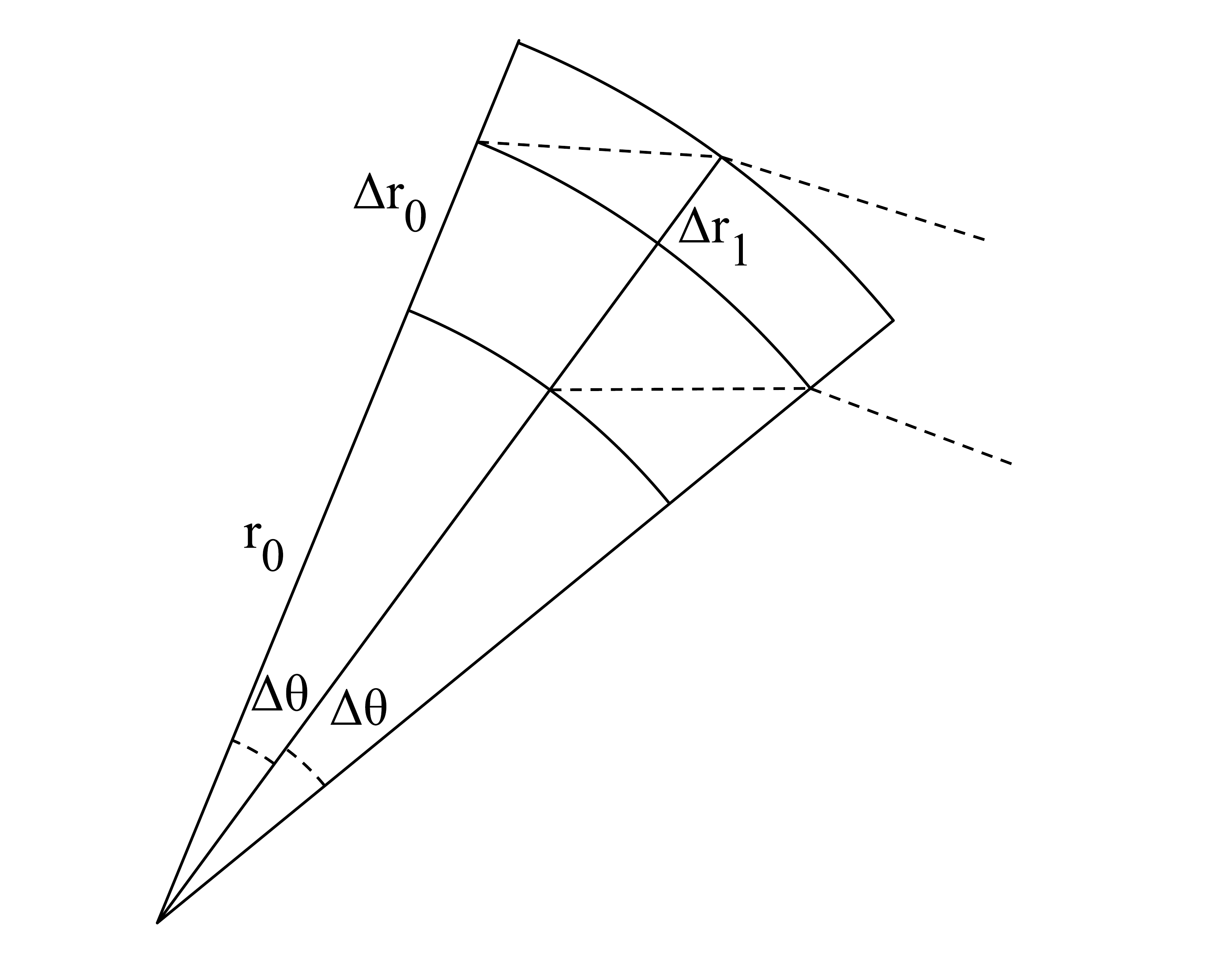}
\caption{Design of a sector for circular moir{\'e} pattern. $\Delta r_{1}$ is chosen such that $r_{0} \Delta \theta \Delta r_{0}=r_{1} \Delta \theta \Delta r_{1}$. } \label{sector}
\end{figure}
    \begin{figure}[t]
            \centering
            \includegraphics[width=3.4in]{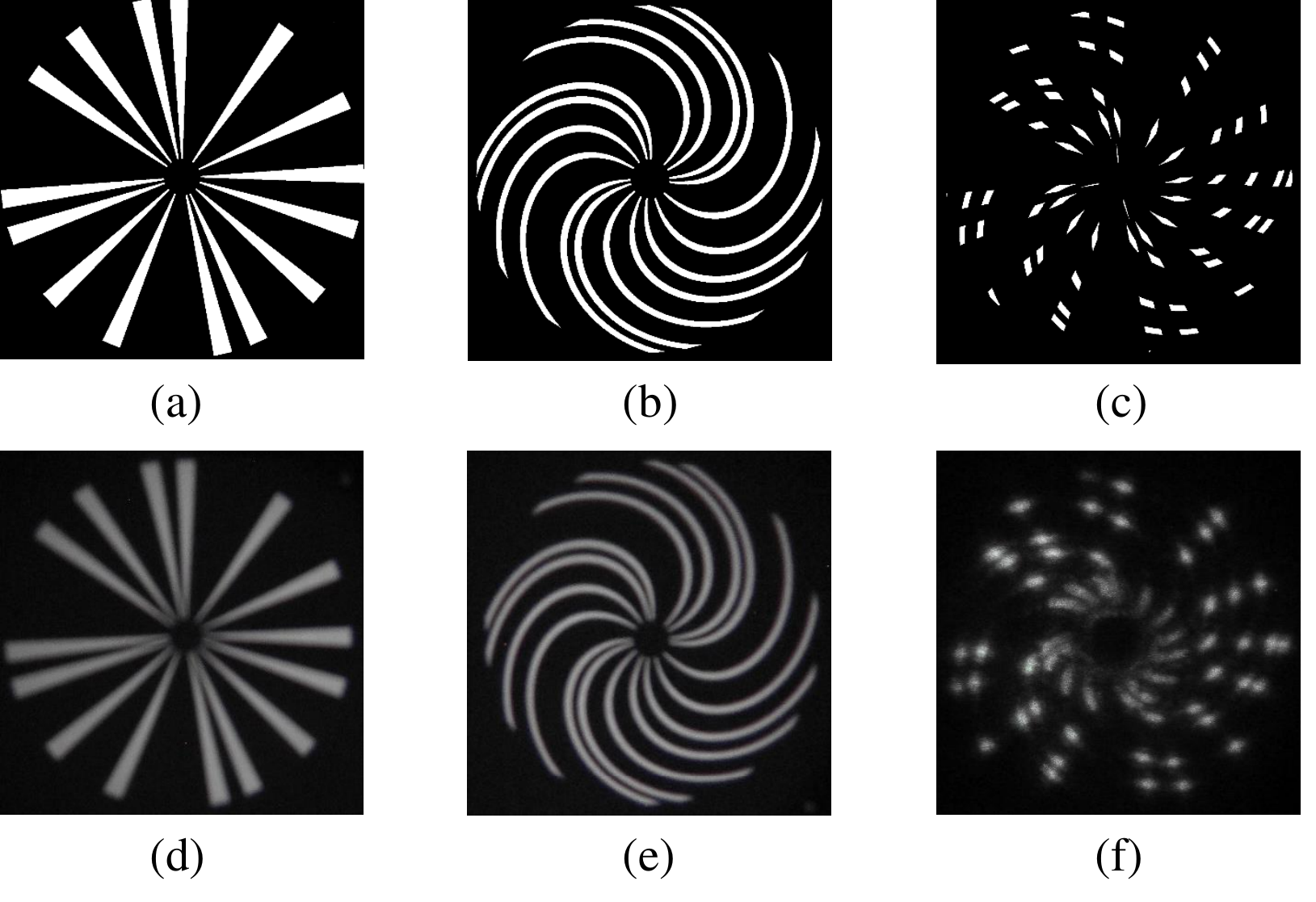}
            \caption {Images of first, second, and output patterns obtained from simulation ((a), (b), and (c), respectively) and experiment ((d), (e), and (f), respectively) in the case of exact matching. }
                        \vspace{0in}\label{Circular}
       \end{figure}
  Our proposed approach is based on encoding the strings into circular images. In this method, instead of rectangular stripes, narrow sectors are applied, as shown in Fig. \ref{Circular}(a). To realize the shifted versions of another string, we use curved pattern, as depicted in \ref{Circular}(b). Each curved sector in this pattern is designed such that the area of its segments at different radii are approximately equal. Defining $r_{0}$ and $\Delta r_{0}$ as our initial values for the first segment of the curved sector, we have (see Fig. \ref{sector});
\begin{equation}
r_{0} \Delta \theta \Delta r_{0}=(r_{0} +\Delta r_{0})\Delta \theta \Delta r_{1}.
\end{equation} 
From the above equation, $\Delta r_{1}$ is given by;
\begin{equation}
\Delta r_{1}=\frac{r_{0}\Delta r_{0}}{r_{0} +\Delta r_{0}}.
\end{equation}
Furthermore, it is obvious from Fig. \ref{sector} that $r_{1}=r_{0} +\Delta r_{0}$. Hence, we can define the following recursive relation to obtain $\Delta r_{i}$s and $r_{i}$s as;
\begin{align}
\Delta r_{i}&=\frac{r_{i-1}\Delta r_{i-1}}{r_{i-1} +\Delta r_{i-1}},\nonumber\\
r_{i}&=r_{i-1} +\Delta r_{i-1},
\end{align}  
respectively. Fig. \ref{Circular}(c) depicts the simulation and the experimental patterns after overlapping two images of Figs. \ref{Circular}(a) and (b). As can be seen, a bright ring appears at the intersection of matched elements.
\section{Experimental Setup and Results}
\begin{figure}
           \centering
           \includegraphics[trim=0.05in 0in 1cm 0in ,clip,width=3.4in]{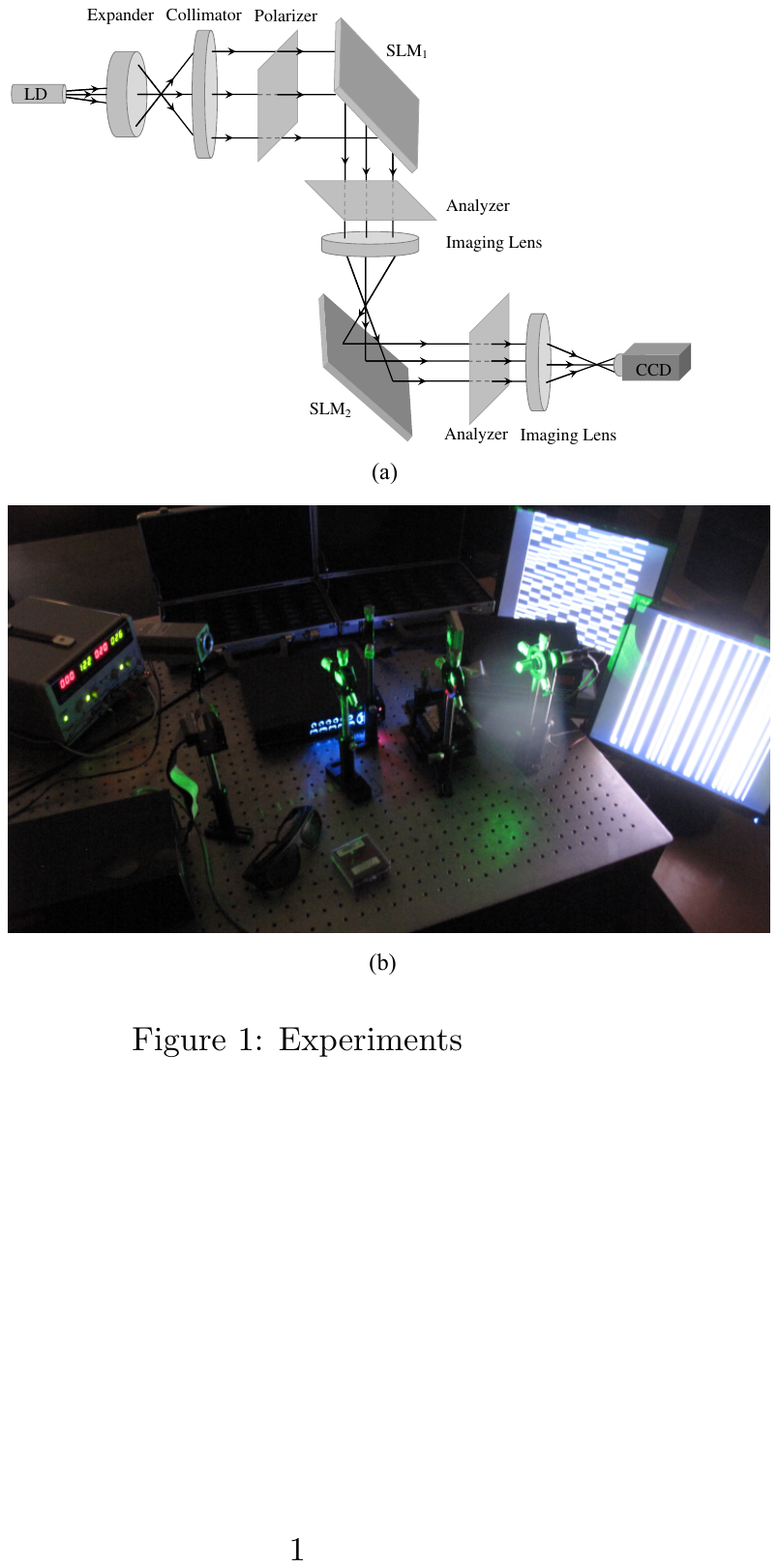}
           \caption {(a) Schematic block diagram and (b) experimental setup for the proposed optical sequence data processing.}
                       \vspace{0in}\label{Setup}
      \end{figure}

In this paper, the optical architecture is implemented by two separate programmable reflective SLMs. The pixel pitch of the liquid-crystal display of each SLM is $20~\rm{ \mu m}$, and the pixels number is $1280\times768$. The larger number of elements it comprises, the less resolution in the output plane we achieve. Further details about equipments can be found in Table \ref{T3}. Fig. \ref{Setup} shows the architecture of multiplier realizing the proposed optical processing method for genome analysis based on spatial coded  technique. The output light of a spatially coherent source such as a laser diode source or a non-coherent source like LED is collimated and then impinges the first SLM containing $S_1$. A linear polarizer in front of the first SLM sets the incoming polarization state. Since the laser emits elliptical polarized light, the intensity is dependent on both the SLM and the first polarizer's states. To remove this ambiguity, the intensity of the light leaving the first polarizer has to be set to be independent of the angle.

 The reflected light from the first SLM meets the second SLM which implements $S_2$. Since horizontal tilt angle is so small ($<5^\circ$) for a reflective SLM, the two SLMs have to be placed at further distance to realize an appropriate setup. Consequently, the high resolution pattern in $\rm{SLM_1}$ would be damaged in that it convolves with free-space Green's function. Using a lens at a distance of twice the focal length ($2f$) between two SLMs makes it possible to have the exact sharp pattern of $S_1$ on $\rm{SLM_2}$. Moreover, fine adjustment of the first polarizer and the analyzer ensures the maximum contrast in the plane of $\rm{SLM_2}$. Each SLM consists of rectangular pixels where each pixel corresponds to the programmed generated binary element. A pixel with binary element ``$1$" allows light to reflect with the same impinging polarization, ideally without any attenuation, corresponding to the white string and a pixel with binary element ``$0$" rotates the incoming light polarization by $90^\circ$ corresponding to the black string.
        \begin{figure}[t]
        	\centering
        	\includegraphics[width=3.4in]{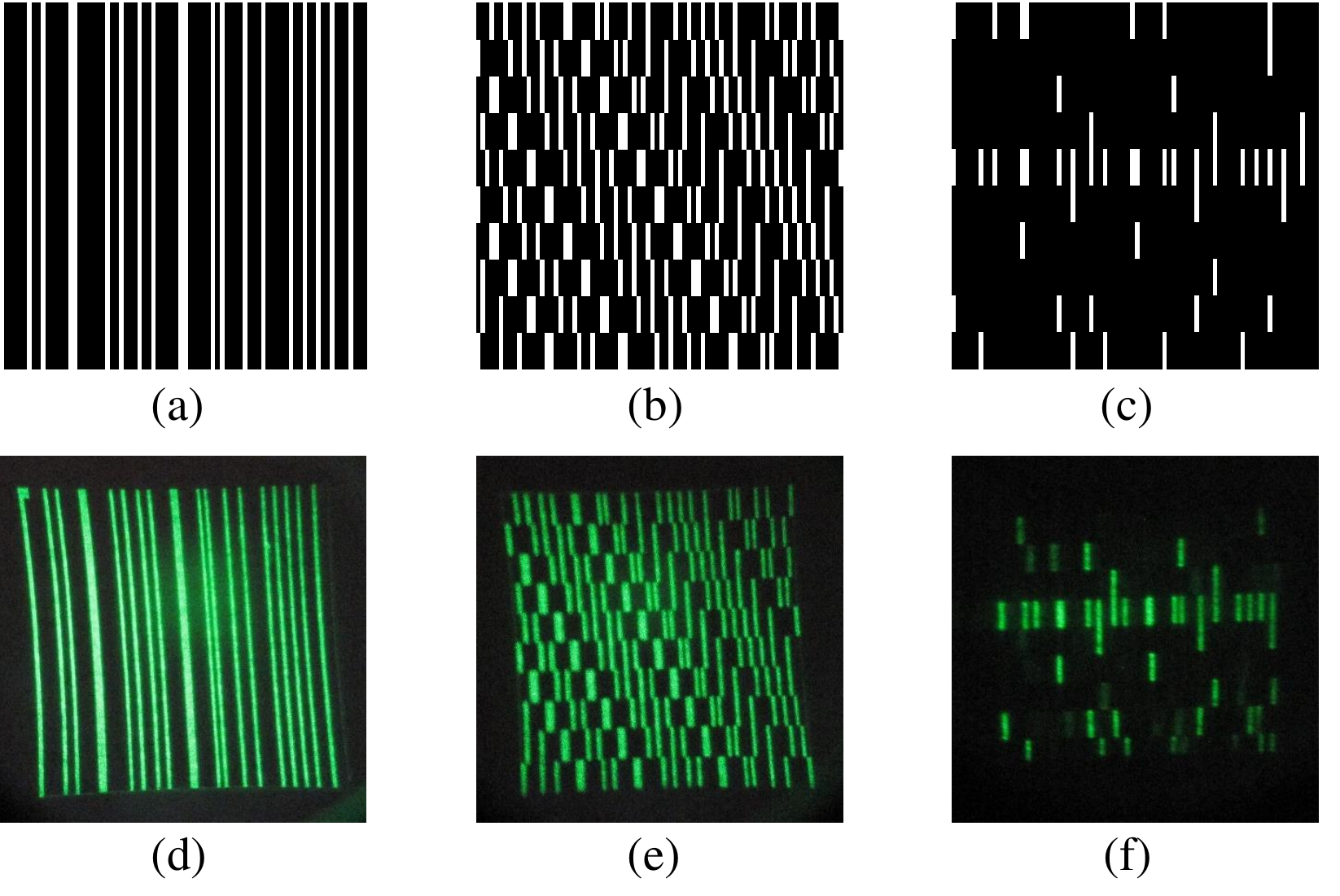}
        	\caption {Image of (a) $S_1$, (b) $S_2$, (c) output pattern achieved via simulation. Photograph of (d) $S_1$, (e) $S_2$, (f) output pattern obtained from experiment.}
        	\vspace{0in}\label{EXP1}
        \end{figure} 
        \begin{figure}[t]
        	\centering
        	\includegraphics[width=3.4in]{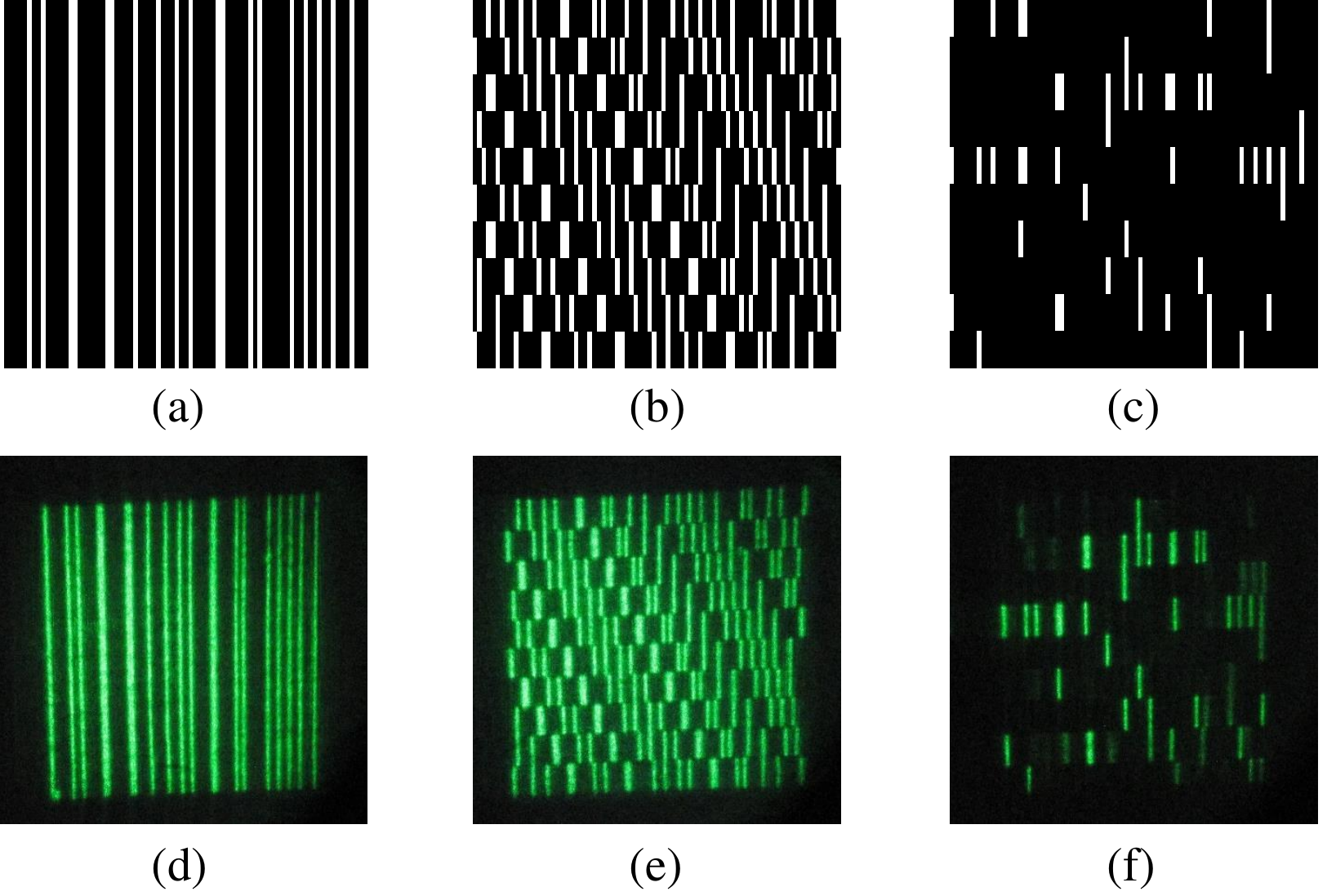}
        	\caption {Image of (a) $S_1$, (b) $S_3$, (c) output pattern achieved via simulation. Photograph of (d) $S_1$, (e) $S_3$, (f) output pattern obtained from experiment.}
        	\vspace{0in}\label{EXP2}
        \end{figure}
        \begin{figure}[t]
        	\centering
        	\includegraphics[width=3.4in]{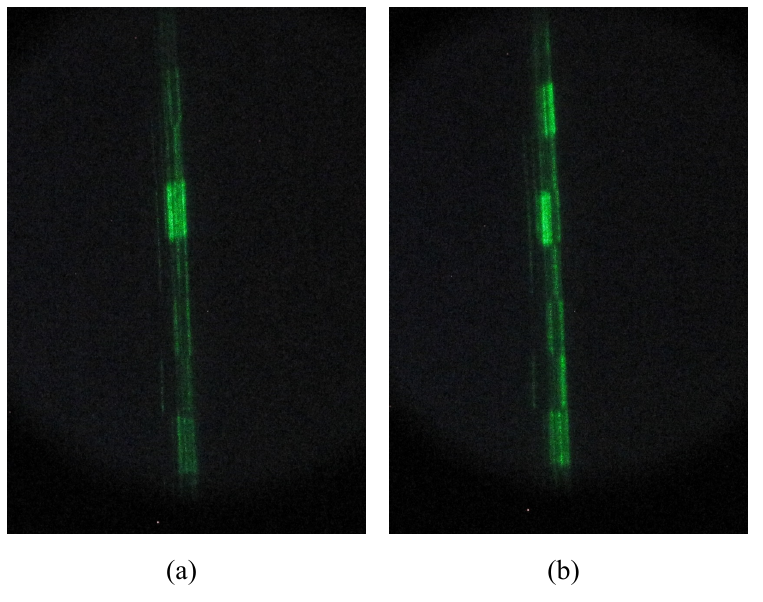}
        	\caption {Transformed output patterns of Figs. \ref{EXP1}(f) and \ref{EXP2}(f) on the display using cylindrical lens.}
        	\vspace{0in}\label{Cylindrical}
        \end{figure}  
        \begin{table}[t]
        	\centering
        	\caption{Optical architecture characterization}
        	\label{T4}
        	\begin{tabular}{c c}
        		
        		Equipment & Description           \\ \hline\hline
        		Reflective SLM      & Holoeye, LC R-720 and LC R-2500\\
        		Biconvex lens & Thorlabs, $f=~75~{\rm and}~100$~mm,~$d=40$~mm\\
        		Laser doide & $1$~mW, Green ($532$~nm), Polarized \\
        		CCD camera & Tevicom\\
        		Achromatic objective lens& Thorlabs, $NA=0.25$\\
        		Cylindrical lens& $f=75$~mm, $h=50.8$~mm, $ l=53 $~mm\\
        		Holder& Standa\\
        		Analyzer and polarizer & Wire grid\\ \hline \hline
        	\end{tabular}
        \end{table}
  A two-dimensional array of photodetectors can be employed to capture the output pattern; then digital processing would be done by a host computer to extract precise matching. Alternatively, analyzing the output pattern can be realized by visual inspection or using a CCD camera.
  
In order to verify our proposed method, we firstly show bar strings alignment between two DNA-simulated sequences; then  the circular one will be demonstrated as our proposed new encoded pattern. One-dimensional strings to be aligned are illustrated in Figs. \ref{EXP1} and \ref{EXP2} in which $S_1$, $S_2$, and $S_3$ were introduced earlier. To more straightforwardly realize string alignment, a cylindrical lens could be employed between the third polarizer and the output display. It is well known that such a lens transforms plane wave to an ultra-thin line. As a result, each horizontal bright line in the output pattern right behind the lens is mapped to a luminous point on the display which enables us to use a simple one-dimensional array of photodetectors to detect the occurrence of exact matching and the number of deleted or inserted elements. Figs. \ref{Cylindrical}(a) and \ref{Cylindrical}(b) respectively illustrate the transformed versions of the output patterns in Figs. \ref{EXP1}(f) and \ref{EXP2}(f) at the focused plane of the cylindrical lens. Additionally, simulated and experimental results for circular patterns presented in Fig. \ref{Circular} are in good agreement.

     \section{Conclusion}
In conclusion, a simple and practical method based on spatially coded moir{\'e} matching technique has been proposed for string alignment processing. Easy interpretation and inherent parallelism with almost real-time processing are the main specifications of our approach which is compatible with digital devices. The processing gain and SNR of the proposed patterns, i.e., bar and circular patterns, have numerically been calculated to show the effectiveness of our method. Moreover, a preprocessing stage which remarkably decreases post-processing time needed for interpretation of output pattern has been introduced. The capability of our proposed method in DNA sequence matching has been shown via simulation. Finally, experimental results verify the performance of the method in genomics processing applications based on optical computing. 




\end{document}